\documentclass[fleqn,usenatbib]{mnras}

\usepackage{newtxtext,newtxmath,comment}

\DeclareRobustCommand{\VAN}[3]{#2}
\let\VANthebibliography\thebibliography
\def\thebibliography{\DeclareRobustCommand{\VAN}[3]{##3}\VANthebibliography}

\usepackage[utf8]{inputenc}
\usepackage[T1]{fontenc}
\usepackage{lipsum,hyperref,subfig,afterpage}
\hypersetup{ 
			colorlinks=true,
			pdfauthor={Lamman, Claire},
			pdftitle={desiIA-draft},
			citecolor=blue
			}

\usepackage{graphicx}	
\usepackage{amsmath}	
\usepackage{lineno}
\let\oldequation\equation
\let\oldendequation\endequation
\renewenvironment{equation}
  {\linenomathNonumbers\oldequation}
  {\oldendequation\endlinenomath}
\usepackage[dvipsnames]{xcolor}
\usepackage{orcidlink}

\title[Redshift-dependent RSD bias from Intrinsic Alignment]{Redshift-dependent RSD bias from Intrinsic Alignment with DESI Year 1 Spectra}

\author[Claire Lamman]{
Claire Lamman \orcidlink{0000-0002-6731-9329},$^{1}$\thanks{E-mail: claire.lamman@cfa.harvard.edu}
  Daniel Eisenstein,$^{1}$     
  Jessica Nicole Aguilar,$^{2}$
  Steven Ahlen \orcidlink{0000-0001-6098-7247},$^{3}$
  David Brooks,$^{4}$\newauthor
  Todd Claybaugh,$^{2}$
  Axel de la Macorra \orcidlink{0000-0002-1769-1640},$^{5}$
  Arjun Dey \orcidlink{0000-0002-4928-4003},$^{6}$
  Biprateep Dey \orcidlink{0000-0002-5665-7912},$^{7}$
  Peter Doel,$^{4}$
  Simone Ferraro \orcidlink{0000-0003-4992-7854},$^{2,8}$\newauthor
  Andreu Font-Ribera \orcidlink{0000-0002-3033-7312},$^{9}$
  Jaime E. Forero-Romero \orcidlink{0000-0002-2890-3725},$^{11,12}$
  Satya Gontcho A Gontcho \orcidlink{0000-0003-3142-233X},$^{2}$
  Julien Guy,$^{2}$\newauthor
  Robert Kehoe,$^{13}$
  Anthony Kremin \orcidlink{0000-0001-6356-7424},$^{2}$
  Laurent Le Guillou \orcidlink{0000-0001-7178-8868},$^{14}$
  Michael Levi \orcidlink{0000-0003-1887-1018},$^{2}$
  Marc Manera \orcidlink{0000-0003-4962-8934},$^{9}$\newauthor
  Ramon Miquel,$^{9,15}$
  Jeffrey A. Newman \orcidlink{0000-0001-8684-2222},$^{7}$
  Jundan Nie \orcidlink{0000-0001-6590-8122},$^{16}$
  Nathalie Palanque-Delabrouille \orcidlink{0000-0003-3188-784X},$^{17}$\newauthor
  Francisco Prada \orcidlink{0000-0001-7145-8674},$^{18}$
  Mehdi Rezaie \orcidlink{0000-0001-5589-7116},$^{19}$
  Graziano Rossi,$^{20}$
  Eusebio Sanchez \orcidlink{0000-0002-9646-8198},$^{21}$
  Michael Schubnell,$^{22}$\newauthor
  Seo Hee-Jong \orcidlink{0000-0002-6588-3508},$^{23}$
  Gregory Tarlé \orcidlink{0000-0003-1704-0781},$^{22}$
  Benjamin Alan Weaver,$^{6}$
  Zhimin Zhou \orcidlink{0000-0002-4135-0977}$^{24}$
  \\ \\
  $^{1}$Center for Astrophysics $|$ Harvard \& Smithsonian, 60 Garden Street, Cambridge, MA 02138, USA\\
  $^{2}$Lawrence Berkeley National Laboratory, 1 Cyclotron Road, Berkeley, CA 94720, USA\\
  $^{3}$Physics Dept., Boston University, 590 Commonwealth Avenue, Boston, MA 02215, USA\\
  $^{4}$Department of Physics \& Astronomy, University College London, Gower Street, London, WC1E 6BT, UK\\
  $^{5}$Instituto de F\'{\i}sica, Universidad Nacional Aut\'{o}noma de M\'{e}xico,  Cd. de M\'{e}xico  C.P. 04510,  M\'{e}xico\\
  $^{6}$NSF's NOIRLab, 950 N. Cherry Ave., Tucson, AZ 85719, USA\\
  $^{7}$Department of Physics \& Astronomy and Pittsburgh Particle Physics, Astrophysics, and Cosmology Center (PITT PACC), University of Pittsburgh, 3941 \\O'Hara Street, Pittsburgh, PA 15260, USA\\
  $^{8}$University of California, Berkeley, 110 Sproul Hall \#5800 Berkeley, CA 94720, USA\\
  $^{9}$Institut de F\'{i}sica d'Altes Energies (IFAE), The Barcelona Institute of Science and Technology, Campus UAB, 08193 Bellaterra Barcelona, Spain\\
  $^{10}$Department of Physics \& Astronomy, University College London, Gower Street, London, WC1E 6BT, UK\\
  $^{11}$Departamento de F\'isica, Universidad de los Andes, Cra. 1 No. 18A-10, Edificio Ip, CP 111711, Bogot\'a, Colombia\\
  $^{12}$Observatorio Astron\'omico, Universidad de los Andes, Cra. 1 No. 18A-10, Edificio H, CP 111711 Bogot\'a, Colombia\\
  $^{13}$Department of Physics, Southern Methodist University, 3215 Daniel Avenue, Dallas, TX 75275, USA\\
  $^{14}$Sorbonne Universit\'{e}, CNRS/IN2P3, Laboratoire de Physique Nucl\'{e}aire et de Hautes Energies (LPNHE), FR-75005 Paris, France\\
  $^{15}$Instituci\'{o} Catalana de Recerca i Estudis Avan\c{c}ats, Passeig de Llu\'{\i}s Companys, 23, 08010 Barcelona, Spain\\
   $^{16}$National Astronomical Observatories, Chinese Academy of Sciences, A20 Datun Rd., Chaoyang District, Beijing, 100012, P.R. China\\
  $^{17}$IRFU, CEA, Universit\'{e} Paris-Saclay, F-91191 Gif-sur-Yvette, France\\
  $^{18}$Instituto de Astrof\'{i}sica de Andaluc\'{i}a (CSIC), Glorieta de la Astronom\'{i}a, s/n, E-18008 Granada, Spain\\
  $^{19}$Department of Physics, Kansas State University, 116 Cardwell Hall, Manhattan, KS 66506, USA\\
  $^{20}$Department of Physics and Astronomy, Sejong University, Seoul, 143-747, Korea\\
  $^{21}$CIEMAT, Avenida Complutense 40, E-28040 Madrid, Spain\\
  $^{22}$University of Michigan, Ann Arbor, MI 48109, USA\\
  $^{23}$Department of Physics \& Astronomy, Ohio University, Athens, OH 45701, USA\\
  $^{24}$National Astronomical Observatories, Chinese Academy of Sciences, A20 Datun Rd., Chaoyang District, Beijing, 100012, P.R. China
}

\date{Accepted XXX. Received YYY; in original form ZZZ}

\pubyear{2023}

\begin{document}
\label{firstpage}
\pagerange{\pageref{firstpage}--\pageref{lastpage}}
\maketitle

\begin{abstract}
 We estimate the redshift-dependent, anisotropic clustering signal in the Dark Energy Spectroscopic Instrument (DESI) Year 1 Survey created by tidal alignments of Luminous Red Galaxies (LRGs) and a selection-induced galaxy orientation bias. To this end, we measured the correlation between LRG shapes and the tidal field with DESI's Year 1 redshifts, as traced by LRGs and Emission-Line Galaxies (ELGs). We also estimate the galaxy orientation bias of LRGs caused by DESI's aperture-based selection, and find it to increase by a factor of seven between redshifts $0.4-1.1$ due to redder, fainter galaxies falling closer to DESI's imaging selection cuts. These effects combine to dampen measurements of the quadrupole of the correlation function ($\xi_2$) caused by structure growth on scales of 10-80 $h^{-1}$Mpc by about 0.15\% for low redshifts (0.4$<$z$<$0.6) and 0.8\% for high (0.8$<$z$<$1.1), a significant fraction of DESI's error budget. We provide estimates of the $\xi_2$ signal created by intrinsic alignments that can be used to correct this effect, which is necessary to meet DESI's forecasted precision on measuring the growth rate of structure. While imaging quality varies across DESI's footprint, we find no significant difference in this effect between imaging regions in the Legacy Imaging Survey.
\end{abstract}

\begin{keywords}
methods: data analysis --cosmology: observations -- large-scale structure of Universe -- -- cosmology: dark energy
\vspace{-.4in}
\end{keywords}



\section{Introduction}

Measuring the growth of large-scale structure in the Universe informs us about the components that drive it: gravity and dark energy. The main observable used to measure this evolution on large scales is the Kaiser effect \citep{kaiser_clustering_1987}. As structure grows, matter falls towards dense regions. This increases the recessional velocity of matter between us and a dense region, while decreasing it for matter falling in from the other side. The result is a ``squashing" effect in redshift-space. This is the dominant source of redshift-space distortions (RSD) on scales larger than clusters (around $10h^{-1}$Mpc). Clustering is quantified using the correlation function. This can be expressed as a series of spherical harmonics, of which the quadrupole $\xi_2$ describes the anisotropic clustering that arises from RSD. On large scales, the growth rate of structure is linearly related to $\xi_2$. This makes RSD a powerful test of cosmological parameters and measuring it is one of the two main science goals of the the Dark Energy Spectroscopic Instrument (DESI). 

DESI is in the midst of a 5-year survey, measuring spectra of over 40 million galaxies within 16,000 deg$^2$ of the sky. The instrument is installed on the 4-meter Mayall telescope and can gather tens of thousands of extra-galactic spectra in one night with its focal plane, which is comprised of 5000 individually controlled robots to position fibers onto galaxies \citep{levi_desi_2013, desi_collaboration_desi_2016, desi_collaboration_desi_2016-1, desi_collaboration_overview_2022}. 

DESI is forecasting a 0.4-0.7\% measurement of the growth rate of structure, $f\sigma_8$. This is measured through the comparison of anisotropic and isotropic clustering, $\xi_2$ and $\xi_0$. It is more difficult to obtain high precision on $\xi_2$ than $\xi_0$, so errors on $f\sigma_8$ are dominated by $\xi_2$. To meet DESI's science goals, it's imperative to measure $\xi_2$ to a precision of at least 0.4-0.7\%. A subtle effect that could take up a significant fraction of this error budget is the bias in $\xi_2$ due to a combination of two effects: Intrinsic Alignment (IA) and a selection-induced orientation bias \citep{hirata_tidal_2009}.\par

IA refers to physical correlations between galaxy shapes and with galaxy shapes to the underlying density. 
See \cite{lamman_ia_2023} for a pedagogical guide to IA and \cite{joachimi_galaxy_2015}, \cite{troxel_intrinsic_2015} for detailed reviews. It is historically measured as a contaminant of weak lensing, but IA in upcoming surveys may provide novel constraints on galaxy formation and cosmology \citep{chisari_cosmological_2013, okumura_first_2023, kurita_constraints_2023, xu_evidence_2023}. For DESI, IA also needs to be understood as a bias in measurements of anisotropic clustering.\par

This effect arises from the extent to which galaxy shapes are correlated with the underlying tidal field. The primary axis of Luminous Red Galaxies (LRGs) tend to be aligned along strands of density and point towards denser regions. This creates a clustering bias when combined with DESI's aperture-based target selection. An elliptical galaxy with its primary axis pointed at the observer will have a more concentrated light profile on the sky and a higher fraction of its light will fall within the aperture. This makes DESI more likely to observe galaxies which lie in density filaments that are parallel to the line of sight (LOS). For a visualization of this effect, see Figure 1 in \cite{lamman_intrinsic_2023}. Studies have explored the effects of orientation-dependent selection in Sloan Digital Sky Survey (SDSS) catalogs with differing results \citep{martens_radial_2018, obuljen_detection_2020, singh_fundamental_2021}. We expect this effect to be more pronounced with DESI, which has a smaller fiber aperture of 1.5 arcsec in diameter, as opposed to SDSS' 3 arcsec aperture.\par

A total-magnitude selection would remove this bias from DESI, but spectroscopic success is highly dependent on the surface brightness of an object. Especially for a survey which prioritizes speed, there will be a surface-brightness dependence on the sample which is easier to impose explicitly as a target cut \citep{zhou_target_2022}. This selection-induced bias in galaxy orientation likely also affects DESI's Emission-Line Galaxy (ELG) sample. However, ELGs are not the primary tracers DESI uses for measuring large-scale structure and they display weaker alignments. While predicted by simulations, there is currently no observed shape alignment in ELG-like (spiral) galaxies \citep{samuroff_dark_2019, johnston_kidsgama_2019, samuroff_dark_2023}.\par

Measuring IA for the purpose of predicting an RSD bias has a few differences from IA measured in the context of weak lensing, which requires very precise shape measurements with well controlled systematic effects to measure gravitational shear. 
We only require shape measurements which are more precise than intrinsic shape variation. This is the case with LRGs in DESI's Legacy Imaging Survey, which are relatively large and bright. Therefore it is more valuable for us to use the full redshift sample available than limit to a region which overlaps with a deeper imaging survey, as will be done with other DESI IA measurements (Lange et al. in prep). 
Also, since this study uses spectroscopic redshifts, we can sufficiently isolate pairs with low separation along the LOS to the degree that we are unconcerned about contamination from weak lensing or across redshift bins. \par


\cite{lamman_intrinsic_2023} used photometric redshifts from DESI's imaging catalog to estimate that this effect could lower DESI's measurement of $\xi_2$ by about 0.5\% for LRGs. In this work, we use DESI's Year-one spectra (DESI Collaboration, in prep) to produce estimates which can be used to correct DESI's RSD measurements. We measure the tidal alignment of LRGs as traced by LRGs and ELGs, assess the impacts of imaging on the IA measurement, and estimate the redshift dependence of the selection-induced shape polarization. We report the resulting redshift-dependent bias for DESI's Year one RSD results and discuss sources of systematic uncertainties.

\section{DESI Catalogs}\label{sec:desi_catalog}

\subsection{Imaging}\label{sec:imaging}
DESI's targets are chosen from DR9 of the Legacy Imaging Survey \citep{dey_overview_2019, myers_target-selection_2023}. This contains imaging of sources in $14,000\deg^2$ of the extra galactic sky from three different telescopes: Mayall $z$-band Legacy Survey at the Mayall telescope at Kitt Peak (MzLS), the DECam Legacy Survey from the Blanco telescope at Cerro Tololo (DECaLS), and the Beijing-Arizona Sky Survey from the BOK telescope at Kitt Peak (BASS) \citep{zou_project_2017}. A region in the South Galactic cap also contains imaging from the DES survey.\par
Their shapes are fit using T\textsc{ractor} \citep{lang_tractor_2016}. After deconvolving with a point-spread function (PSF), shapes are modeled at the pixel level with several light profiles: exponential disk, de Vaucouleurs, Sersic, PSF, and round-exponential. The default shape parameters are chosen based on a modified $\chi^2$ criteria which avoids over-fitting bright targets as round-exponentials. Measuring intrinsic alignment requires shape orientations, so where galaxies are fit as circles (PSF and round-exponentials), we use shape parameters from the best fit between exponential disk and de Vaucouleurs. This will not affect our final results as the DESI target selection does not depend on these derived shape parameters.\par
\begin{figure} 
\begin{center}
    \includegraphics[scale=0.2]{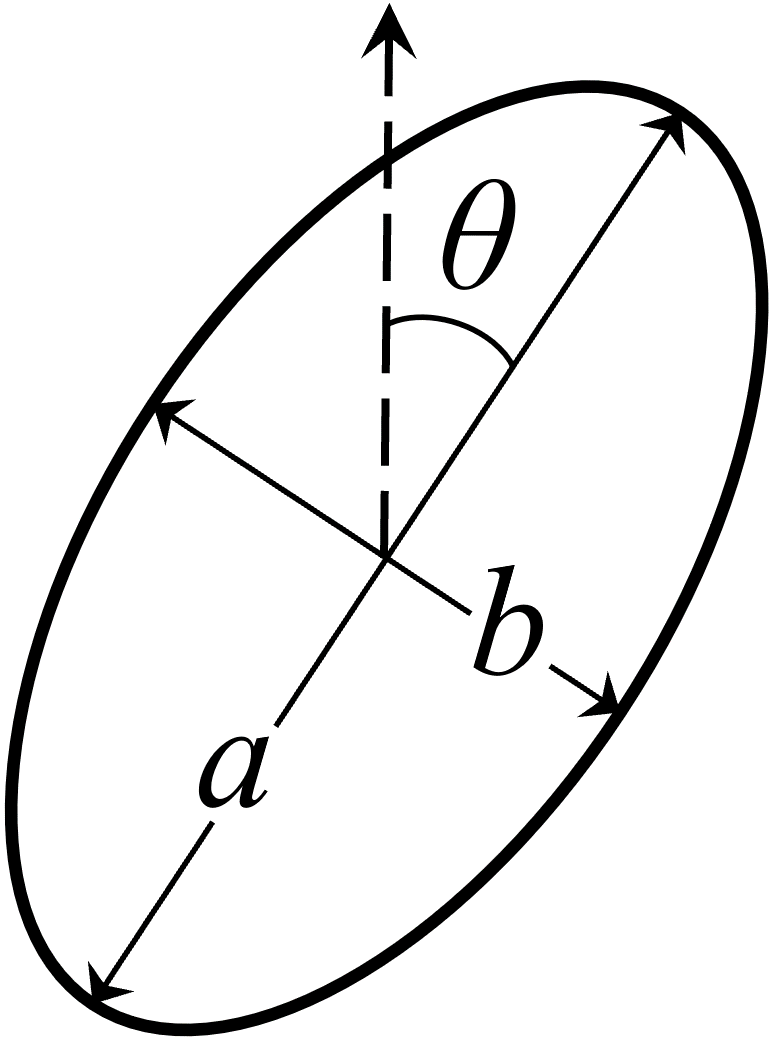}
\end{center}
\caption{The parameters used to describe the shape and orientation of the ellipse created by projecting an elliptical galaxy. Here, the ellipticity is measured relative to North. For our measurement, ellipticity is measured relative to the positions of a tracer sample.}
\label{fig:ellipticity_definition}
\end{figure}

The parameters used to describe each projected galaxy image are its primary axis, $a$, secondary axis, $b$, and orientation $\theta$ (Figure \ref{fig:ellipticity_definition}). This is used to describe the shape of an ellipse relative to some direction using $\epsilon_+$:
\begin{equation}
    \epsilon_+ = \frac{a-b}{a+b}\cos(2\theta)
\end{equation}

DESI's LRG target selection from this catalog includes a cut based on the expected flux which falls within a DESI fiber, which corresponds to a 1.5 arcsec diameter aperture. The $z$-band magnitude within the aperture is limited to $z_{\text{fiber}}<21.61$ in the Northern Galactic Cap and $z_{\text{fiber}}<21.60$ in the Southern Galactic Cap. For more information DESI's target selection, see \cite{raichoor_preliminary_2020, zhou_preliminary_2020, zhou_target_2022}.\par

\begin{figure} 
\begin{center}
    \includegraphics[scale=0.25]{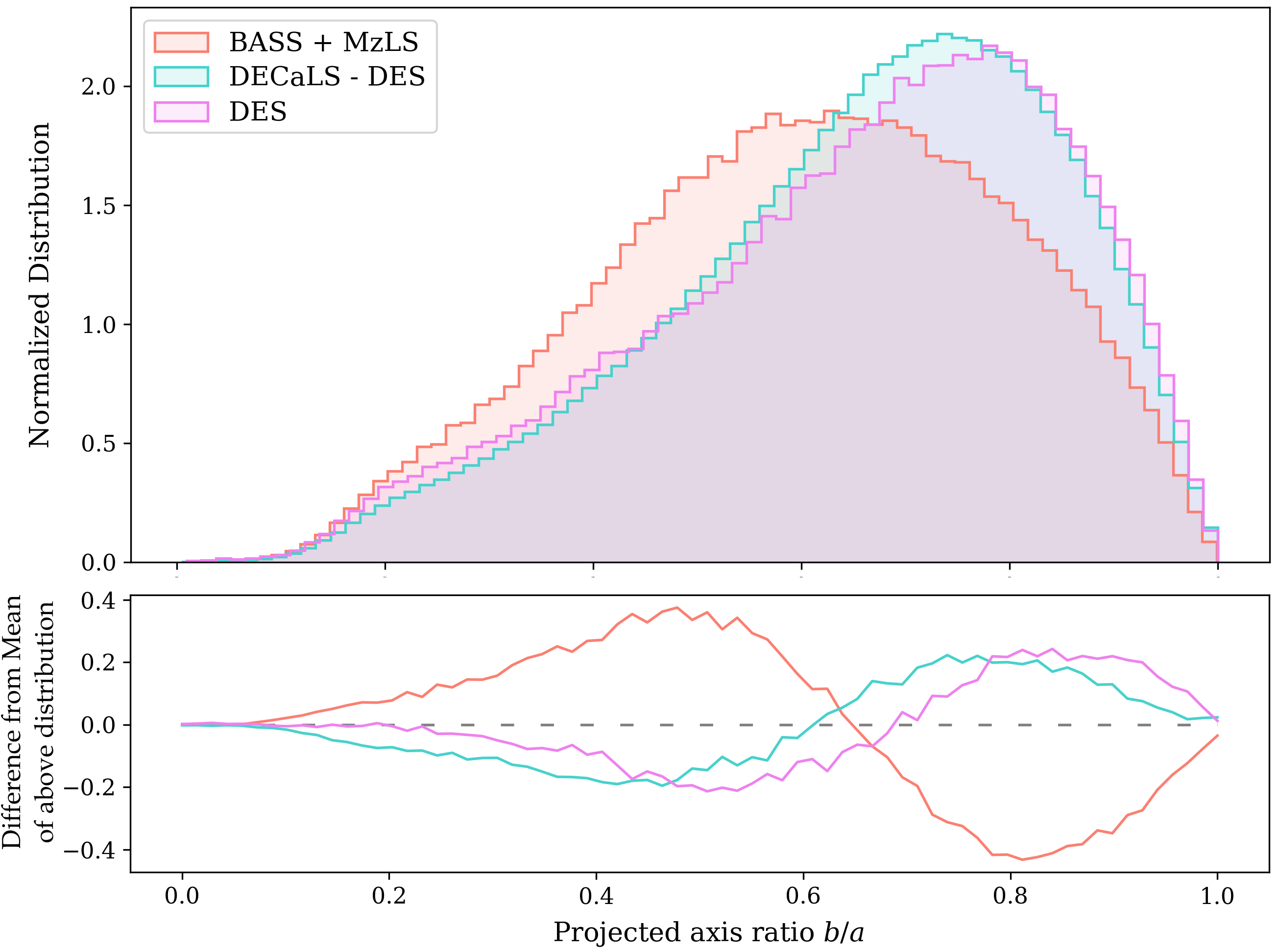}
\end{center}
\caption{The distribution of projected axis ratios for LRGs in three DESI imaging regions: the North Galactic cap, which contains imaging from MzLS and BASS, the portion of the South Galactic cap which contains DECaLS but no DES imaging, and the DES region. Residuals from the mean are plotted below. These reported shapes have been deconvolved with a point spread function to account for imaging conditions. The region with the highest quality imaging, DES, reports the least eccentric shape.}
\label{fig:shape_comparison}
\end{figure}

This shape fitting and target selection is dependent upon imaging quality, which varies across sky regions. To qualify the effect of imaging quality on shape parameters, we separate the LRGs into three sky regions: The MzLS and BASS region, the DECaLS region which does not contain DES imaging, and the DES region. We compare the reported axis ratio, $b/a$, of the reported galaxy shapes in each region in Figure \ref{fig:shape_comparison}. The MzLS and BASS region reports more eccentric LRG shapes than the other regions, and the region with highest-quality imaging, DES, reports the roundest shapes. While this may indicate an over-correcting of the PSF in MzLS and BASS imaging, we measure the IA signal independently in these regions and do not find any significant impact on final results (Section \ref{sec:IA_imaging}).

\subsection{Spectroscopy}\label{section:desi_redshifts}
We used spectra from DESI's internal data release, Iron, which is comprised of data from commissioning through Year 1 of the survey (DESI Collaboration, in prep). It contains spectra of 2.9 million LRGs and 4.0 million ELGs from observations taken during December 14th, 2020 through June 13th 2022 \citep{zhou_target_2022, raichoor_target_2023, guy_spectroscopic_2023}. For approximate redshift distributions of the samples, see Fig 2 of \cite{desi_collaboration_early_2023}. We used Iron's large-scale structure catalog which includes a veto on targets based on hardware and imaging conditions. The catalog contains weights to account for completeness based on the probability that each target was assigned. LRGs also have a weight to account for redshift failures. From this catalog, we ran a basic redshift-quality cut to obtain our final samples of 2.5 million LRGs and 3.1 million ELGs.\par 
The projected correlation functions used to calibrate our measurements (Section \ref{subsec:formalism}) were made with this Year 1 data. Due to internal blinding policies, our estimation of DESI's $\xi_2$ measurements are calibrated with the smaller, publicly available spectroscopic catalog from DESI's Survey Validation \citep{desi_collaboration_early_2023, desi_collaboration_validation_2023, lan_desi_2023}. Our determination of the $\xi_2$ signal which arises from IA is independent of the RSD $\xi_2$ signal.

\section{Intrinsic Alignment Measurement}\label{sec:alignment}

\subsection{IA Formalism}\label{subsec:formalism}

As in \cite{lamman_intrinsic_2023}, we measure the correlation between galaxy shapes and density by averaging the ellipticity of each LRG relative to the separation vector between it and nearby galaxies in the tracer sample\footnote{code available here: \href{https://github.com/cmlamman/ellipse_alignment}{github.com/cmlamman/ellipse\_alignment}}.
\begin{equation}\label{eq:e1}
    \mathcal{E}(r_p) = \langle \epsilon_+(a, b, \theta) \rangle
\end{equation}

\noindent For a given galaxy-tracer pair, $a$ and $b$ are the axis lengths of the galaxy shape and $\theta$ is the orientation of the galaxy relative to the separation vector between it and the tracer. This is measured as a function of the projected separation between them, $r_p$.\par
We limit the separation of pairs along the LOS, $r_\Pi$, to $\pm \Pi_\text{max}=$ 30$h^{-1}$Mpc. This, along with clustering, is taken into account in our model when estimating how far along the LOS the IA measurement is averaged over.\par

For measuring the IA of our full LRG sample, we divide the tracer catalog into 100 sky regions based on right ascension and declination with an equal number of galaxies in each. $\mathcal{E}(r_p)$ is measured independently in each region using its tracers and the full shape catalog, then averaged over every pair. This average included the catalog weights described in Section \ref{section:desi_redshifts} for both the shape and tracer samples. The average of these 100 measurements and standard error is our final measurement. \par


IA is often quantified using a form of correlation functions generalized to include information about galaxy shapes \citep{mandelbaum_detection_2006}. The IA correlation function relating galaxy shapes and density is
\begin{equation}
     \xi_{g+}(r_p, \Pi) = \frac{S_+D - S_+R_D}{R_SR_D}.
\end{equation}
$S_+D$ is the count of data-data pairs weighted by the orientation of shapes, $S_+$, relative to a tracer sample $D$. Measuring this as function of projected separation and averaging over each data pair, $S_+D(r_p) / DD(r_p)$, is equivalent to $\mathcal{E}(r_p)$. $S_+R$ represents the data shapes relative to a random sample, which has an expectation value of 0. $R_SR_D$ is the random-random count. Integrating $\xi_{g+}$ along the LOS direction, $\Pi$, produces the projected IA correlation function:
\begin{equation}
     w_{g+}(r_p) = \int_{-\Pi_\text{max}}^{\Pi_\text{max}} \mathrm{d}\Pi\, \xi_{g+}(r_p, \Pi)
\end{equation}

For predicting the RSD bias that arises from IA, $\mathcal{E}(r_p)$ is the most direct relevant observable. Unlike $w_{g+}$, $\mathcal{E}(r_p)$ is normalized by data pairs, not randoms. $DD$ can be expressed as
\begin{equation}
    \begin{array}{l}
    DD(r_p) = RR\int_{-\Pi_\text{max}}^{\Pi_\text{max}} d\Pi \frac{DD(r_p, \Pi)}{RR} \\
    = RR\int_{-\Pi_\text{max}}^{\Pi_\text{max}} d\Pi (1 + \xi(r_p, \Pi)) = RR(2\Pi_\text{max} + w_p(r_p)).
  \end{array}
\end{equation}
Here $\xi$ and $w_p$ are the typical correlation function and projected correlation function, as opposed to those weighted by shape alignments. $w_{g+}$ can be expressed as
\begin{equation}
    w_{g+}(r_p) = \frac{1}{RR}\int_{-\Pi_\text{max}}^{\Pi_\text{max}} d\Pi S_+D(r_p, \Pi) 
    = \frac{S_+D(r_p)}{RR}.
\end{equation}
Therefore, a given $w_{g+}$ and $\mathcal{E}$ made with the same $\Pi_{\rm max}$ and same clustering $w_p$ are related as
\begin{equation}
    w_{g+}(r_p)= (2\Pi + w_p(r_p))\mathcal{E}(r_p) = L\mathcal{E}(r_p) 
\end{equation}
Here we have introduced $L$, which can be understood as the effective LOS distance that $\mathcal{E}$ is measured over, adjusted to account for clustering which decreases the average LOS-separation of pairs. $L$ is included in our final model of the RSD bias (Section \ref{sec:result})\footnote{$L$ here is equivalent to $L_\text{eff}$ in \cite{lamman_intrinsic_2023}}. While $L\mathcal{E}$ is functionally equivalent to $w_{g+}$, we notate $L$ and $\mathcal{E}$ separately to be explicit about how the quantity was estimated.\par

\begin{figure} 
\includegraphics[width=.48\textwidth]{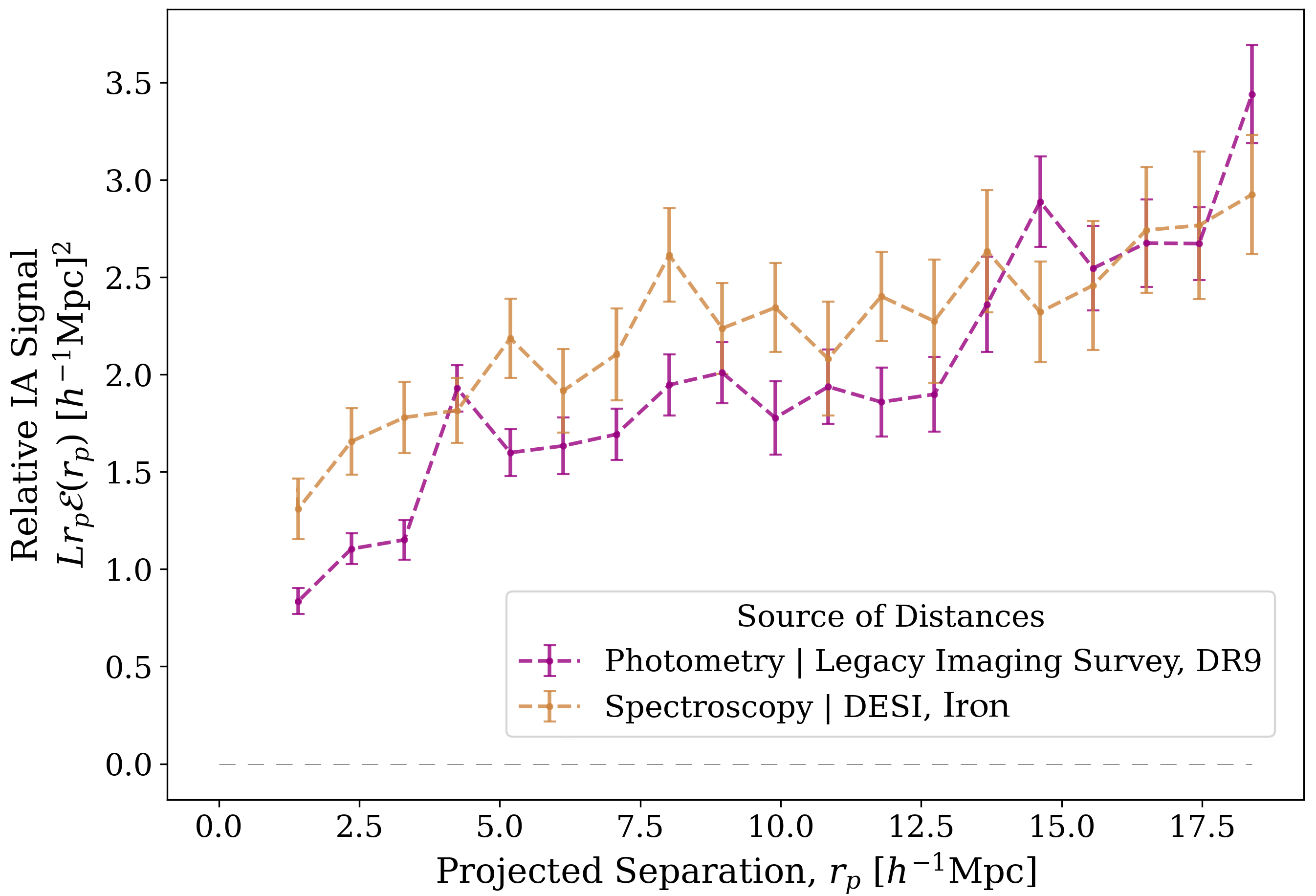}
\caption{The IA signal of LRGs over the entire redshift range, 0.4 $<$ z $<$ 1.1, compared to the estimate made in \citep{lamman_intrinsic_2023} with photometric distances. The photometric estimate was made with 17.5 million galaxies, compared to 2.5 million LRGs for the spectroscopic sample, but necessarily averaged over a larger radial distance. This is adjusted for here, which shows the "relative" IA signal that has been calibrated by the effective radial depth $L$.}
\label{fig:spec_comparison}
\end{figure}

\begin{figure}
\includegraphics[width=.48\textwidth]{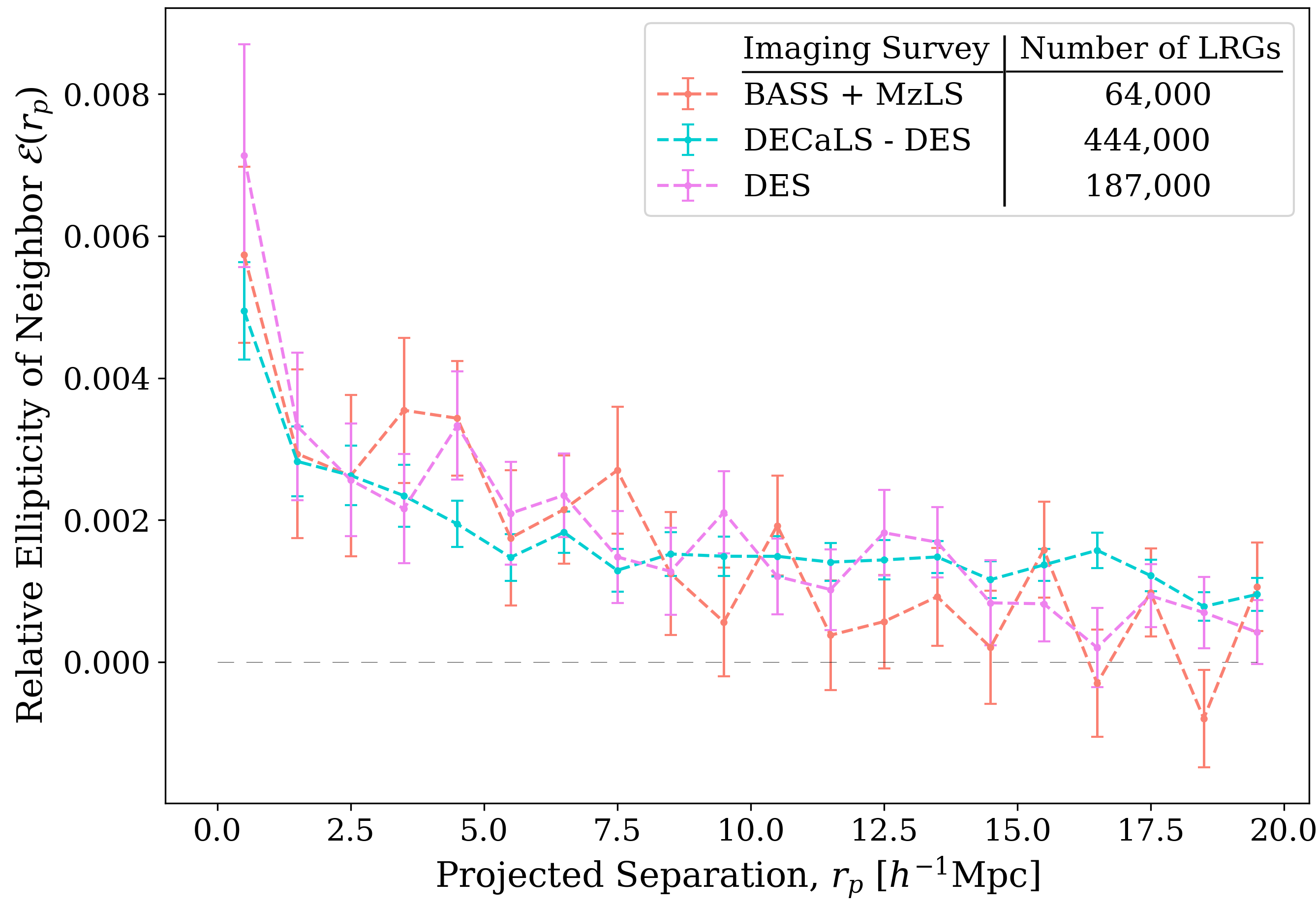}
\caption{Measurement of the tidal alignment of LRG shapes made independently in areas from three regions of DESI's Legacy Imaging Survey described in Section \ref{section:desi_redshifts}. DES has the highest quality imaging, but there is no significant effect on our total averaged IA signal.}
\label{fig:ia_region}
\end{figure} 

We can compare our spectroscopic IA measurement with a similar one made with photometric data \citep{lamman_intrinsic_2023} by scaling by $L$, as shown in Figure \ref{fig:spec_comparison}. Although made with seven times fewer galaxies, the spectroscopic sample from Iron can be measured in smaller LOS bins and provide us with similar level of precision.\par

Spectroscopic data also allows us to better isolate the sample in radial bins and explore redshift-dependence. To compare our IA signal between samples of different target classes and redshift distributions, $\mathcal{E}(r_p)$ needs to be calibrated by $L$ as well as the galaxy clustering bias, $b$. 
For bias-independent comparisons, we scale by a relative bias. The bias of a sample $2$ relative to sample $1$ is
\begin{equation}\label{eq:bias}
    b_{\rm rel}=\frac{b_2}{b_1} = \frac{D(z_1)}{D(z_2)}\bigg(\frac{w_{p2}}{w_{p1}}\bigg)^{1/2},
\end{equation}
where $D(z)$ is the linear growth function. \par

Therefore, when comparing IA measurements across samples we use the value $(L/b_{\rm rel})\mathcal{E}(r_p)$. While $L$ is taken into account when estimating the final RSD bias, $b$ does not affect the final result. This is because the amplitude of the power spectrum quadrupole effect arises from the correlation of the galaxy density field and the selection-induced shape polarization, the latter of which is independent of bias. \par
When calculating distances and the growth factor, we assume a flat $\Lambda$CDM cosmology with $\Omega_m=0.286$, $\Omega_\Lambda=0.714$ and $H_0=69.6\text{ km s}^{-1}\text{Mpc}^{-1}$.\par

\subsection{Dependence on Imaging Region}\label{sec:IA_imaging}
The amplitude of IA can strongly depend upon imaging quality and the methods used to estimate shapes. This is in part due to difficulties in accurately modeling imaging processes, and in part due to isophotal twisting \citep{fasano_isophotal_1989}, which causes the outer regions of galaxies to have a stronger alignment signal than the inner regions. This has been measured in BOSS LOWZ, DES, and LSST \citep{singh_intrinsic_2015, zuntz_dark_2018, leonard_measuring_2018, georgiou_dependence_2019, macmahon_intrinsic_2023}. This is a concern for our imaging catalog, as imaging quality and shapes vary across region (Figure \ref{fig:shape_comparison}). A lower signal due to poor imaging will result in an underestimate of our final estimate of the RSD bias.\par

To qualify the impact of imaging quality, we compare our IA signal across the three different imaging regions used in the Legacy Imaging Survey: DES, DECaLS, and MzLS+BASS. Each region has varying survey completeness, so to avoid edge effects we made these measurements in a limited area with the most completeness in each region. The result and size of each sample is shown in Figure \ref{fig:ia_region}. We do not find a significant difference in the signals, which is in part due to measurement noise. Additionally, although the BASS + MzLS region may be over-correcting for the PSF and producing more eccentric shapes, systematic imaging effects are uncorrelated with the tidal field. A small change in ellipticity doesn't propagate as an order-unity error on this signal, which is a very small response to the tidal shear. This may be still be an issue for higher signal-to-noise detections beyond DESI Year 1.\par

As a null test, we reproduced this measurement using the cross-component of shape, $\epsilon_\times = \frac{a-b}{a+b}\sin(2\theta)$ instead of $\epsilon_+$ in Equation \ref{eq:e1}. This was consistent with 0 on all scales.

\subsection{Dependence on Redshift and Tracer Sample}
The redshift dependence of IA is unclear \citep{samuroff_advances_2021, zhou_intrinsic_2023, samuroff_dark_2023}, and cannot be directly observed without accounting for luminosity differences across redshift bins. DESI's LRG sample is designed to have a constant co-moving volume with redshift, which results in more luminous, and therefore more aligned, galaxies in high redshift samples. However, since we are only inferring a systematic bias and not any physical trends, we only require the IA of each sample. The IA RSD bias is proportional to the amplitude of this signal, so if not properly accounted for, it could manifest in DESI's results as a false evolution of the growth rate as measured by the quadrupole of the correlation function. Therefore we separate our LRG tracer sample into three sub-samples based on redshift and measure the correlation of LRG shapes in each. \par
\begin{figure}
\includegraphics[width=.44\textwidth]{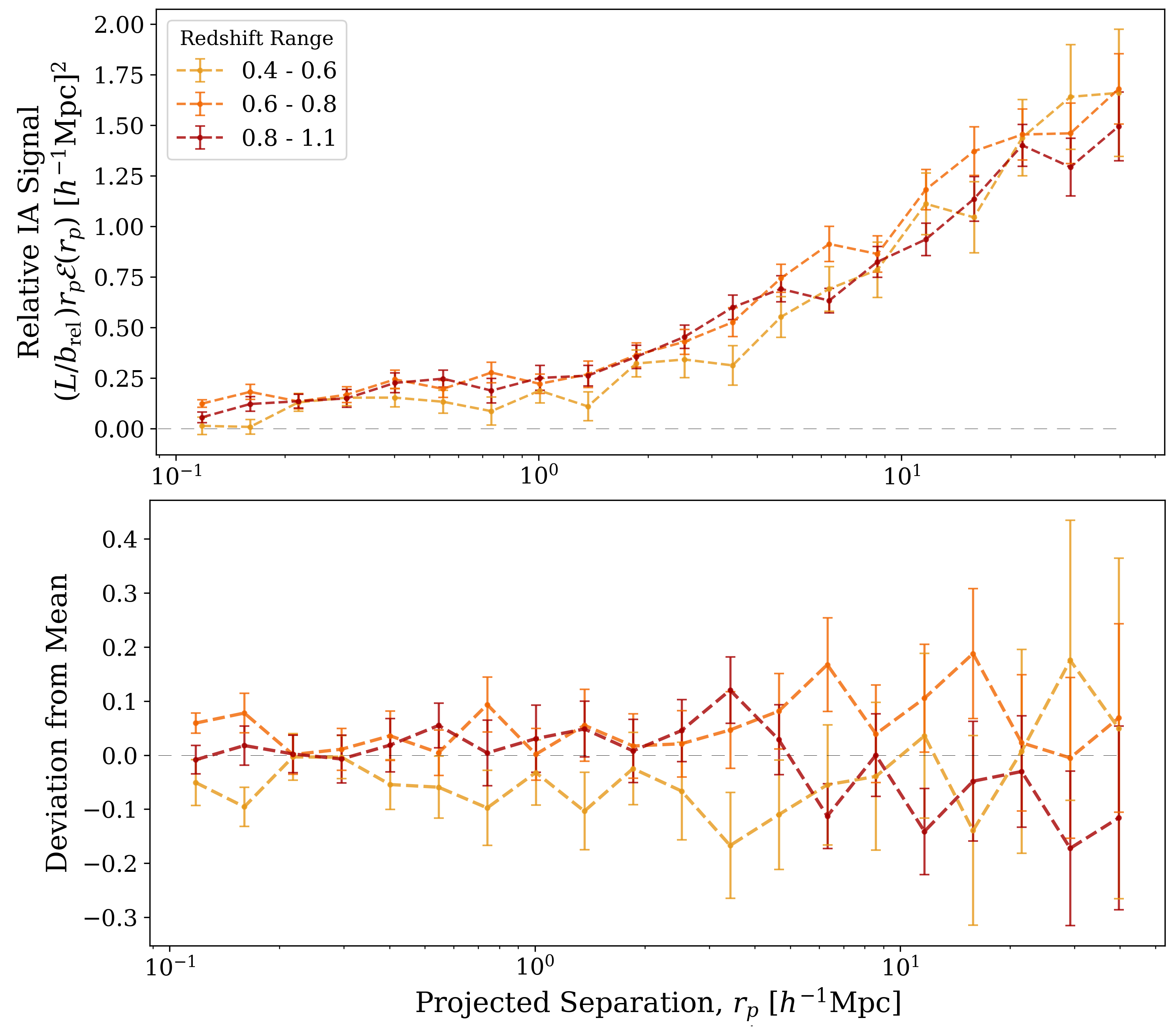}
\caption{Comparison of the intrinsic alignment of LRGs between spectroscopic redshift bins. The y-axis is scaled by the effective depth of the measurement $L$ and the galaxy bias $b_{\rm rel}$, which here is defined as $b_{\rm rel}(z=0.7) = 1$. These were calculated using the projected correlation function from DESI's Year one data. Errors here only include the statistical difference of the signal between sky regions and not from $b$ or $L$. Nearby galaxies broadly display a weaker alignment, though here we have not accounted for luminosity differences across samples.}
\label{fig:ia_redshift}
\end{figure}

The samples are plotted in Figure \ref{fig:ia_redshift} and displayed in Table \ref{tab:redshift_comparison}. To compare the strength of tidal alignment between redshifts, the signal is adjusted based on the clustering in each sample, as described in Section \ref{subsec:formalism}. As expected, we find the weakest signal for nearby galaxies (0.4 $<$ z $<$ 0.6).\par

\begin{figure}
\includegraphics[width=.45\textwidth]{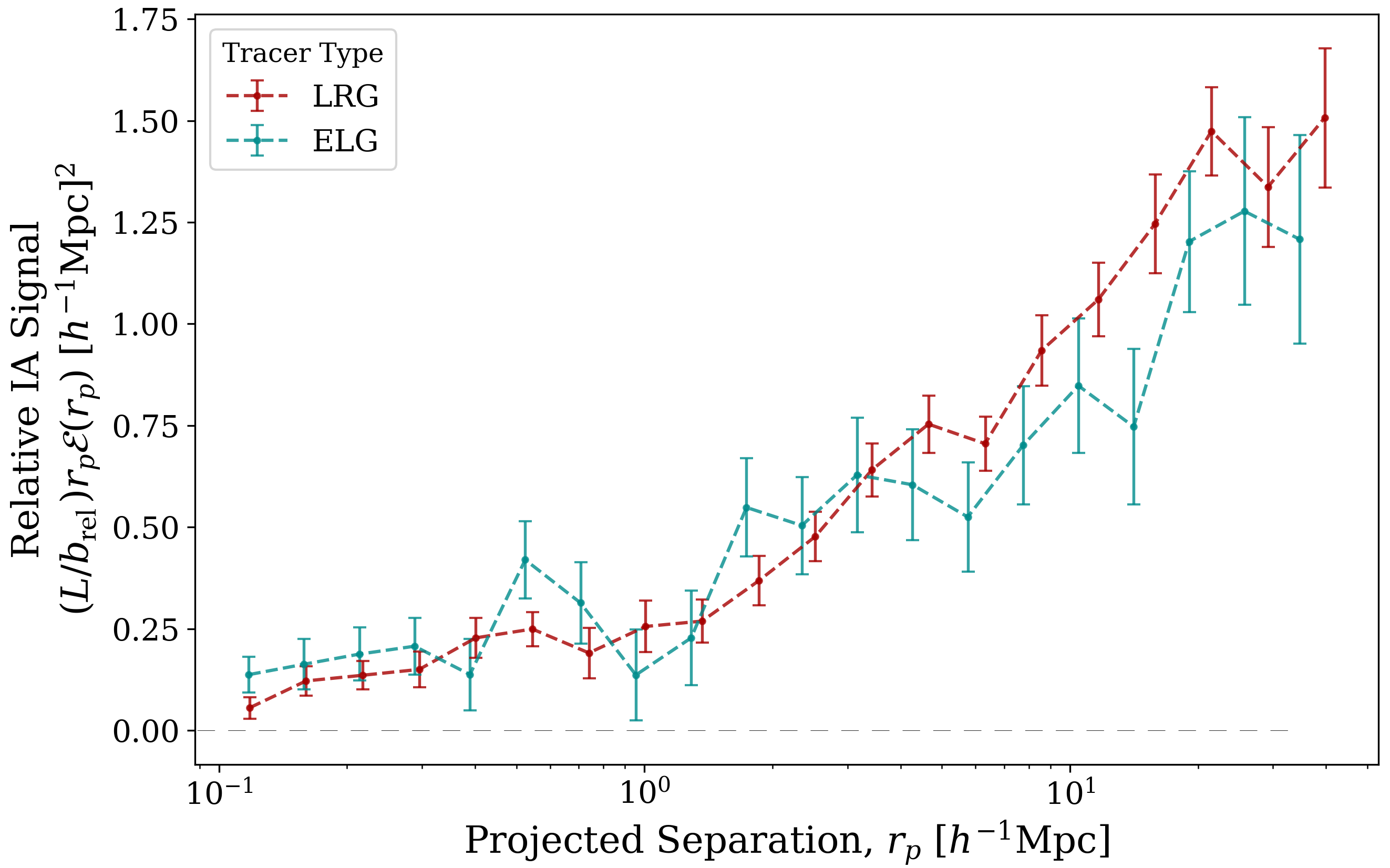}
\caption{Correlation between LRG shapes and the underlying galaxy density, as traced by both LRGs and ELGs. These samples are both in the redshift range 0.8 $<$ z $<$ 1.1. For comparison, this IA signal is scaled by the samples' clustering, as described in \ref{subsec:formalism}.}
\label{fig:ia_tracers}
\end{figure}

\begin{figure}
\begin{center}
\includegraphics[width=.4\textwidth]{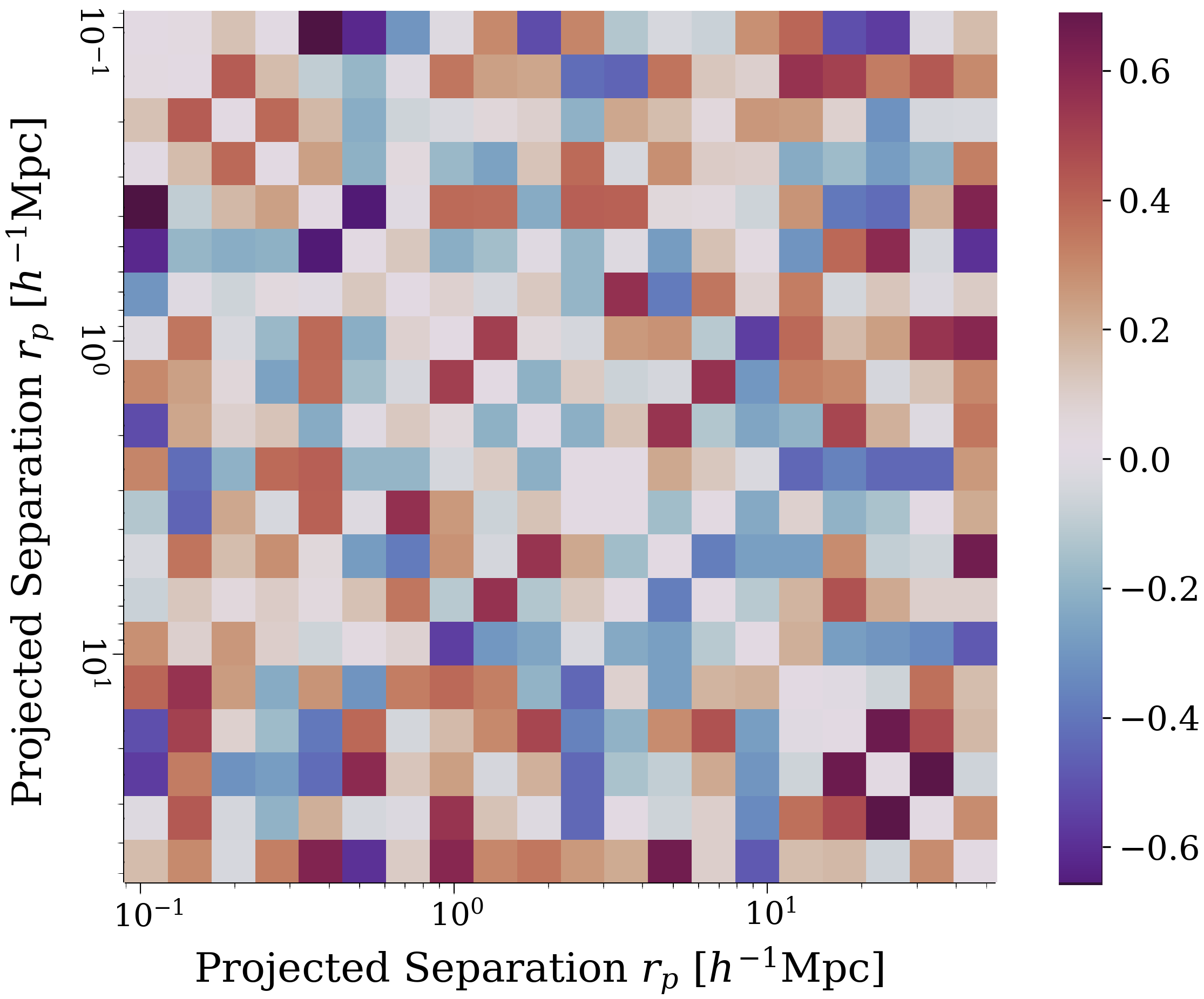}
\end{center}
\caption{The reduced covariance matrix of $\mathcal{E}$ between bins of transverse separation for our IA measurement with LRG tracers across the full redshift range. The identity matrix has been subtracted from this plot. This demonstrates that there is no evidence for significant correlations between the measurements of $\mathcal{E}$ in each bin of projected separation.}
\label{fig:ia_cov}
\end{figure}

We also measured the alignment of LRGs to the tidal field as traced by ELGs, as opposed to the same LRG sample (Figure \ref{fig:ia_tracers}). In the overlapping redshift range of the LRG and ELG samples, $0.8<z<1.1$, we find a similar IA signal once both samples are adjusted for clustering. Although some regions of DESI's Year one footprint are less complete for ELGs, this is accounted for in the catalog completeness weights described in Section \ref{sec:desi_catalog} and we find no impact of this on our IA measurement.\par

A jackknife covariance matrix for the spectroscopic LRG measurement made over all redshift ranges is shown in Figure \ref{fig:ia_cov} and demonstrates no correlation between the bins of projected separation that the IA measurement $\mathcal{E}$ was made in.\par








\begin{figure*}
\includegraphics[width=\textwidth]{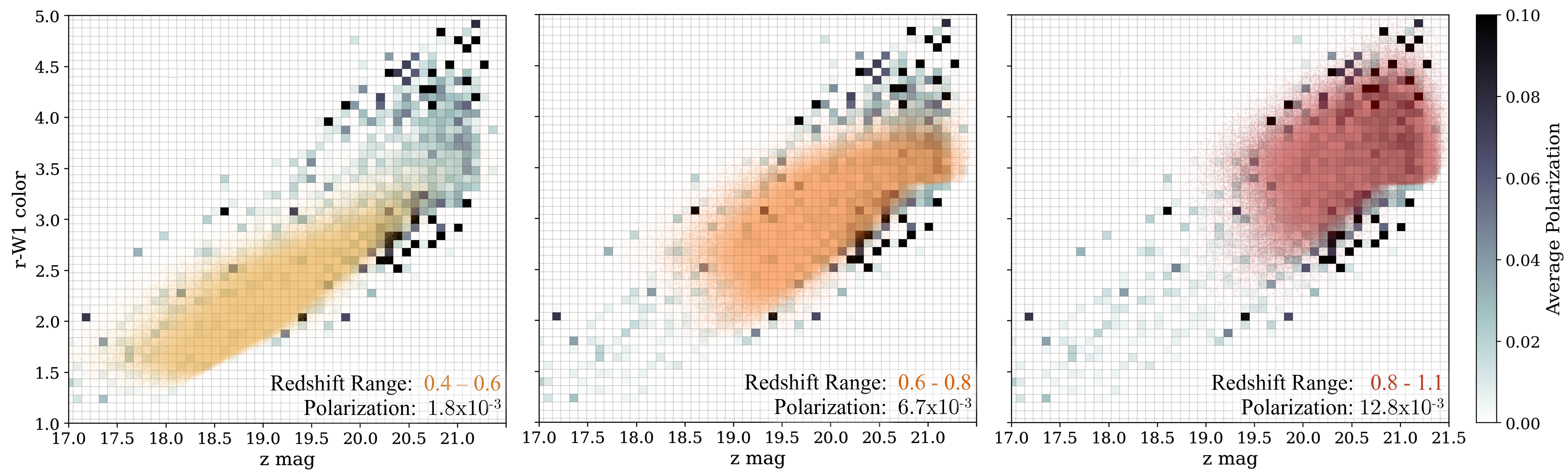}
\caption{Estimates of shape polarization for each of the three redshift bins, or the tendency for these samples to have shapes aligned with the LOS due to an aperture magnitude cut. These are based on the model in \citep{lamman_intrinsic_2023}, shown in grey, which estimates the polarization in bins of color and magnitude for a sample of LRGs without the aperture magnitude cut. The total polarization estimate for each redshift bin is the average of each galaxy's polarization corresponding to its respective color bin. The highest redshift sample contains redder, fainter galaxies which are closer to the survey cuts and therefore more likely to have biased galaxy orientations.}
\label{fig:pol}
\end{figure*}

\section{Selection-Induced Shape Polarization}

The final RSD bias is proportional to both the degree to which galaxies are aligned along the tidal field (IA) and the degree to which galaxies are aligned along the LOS due to target selection, or "shape polarization". For DESI, the latter plays a strong role in the redshift dependence of the RSD bias. Redder, fainter galaxies fall closer to the aperture-magnitude cut that is used to select DESI targets. Therefore their orientation will have a stronger impact on whether or not they are selected; an elongated galaxy aligned with the LOS will have more light concentrated within a sky aperture.\par

\cite{lamman_intrinsic_2023} estimated the shape polarization of DESI LRGs from a parent sample without the aperture-magnitude cut. This was done by generating many 3D light profiles for each galaxy based on the expected distribution of triaxial shapes from \cite{padilla_shapes_2008}. These light profiles were assigned random orientations then put through an aperture-magnitude cut. The average ellipticity of the selected shapes relative to the LOS, $\epsilon_\text{LOS}$, is the selection-induced shape polarization.\par

As this selection is done on aperture magnitudes from an image deconvolved with a point-spread function, the shape selection bias is relatively independent of imaging quality. It matters more to model this effect with imaging that most closely reflects the intrinsic galaxy shape. Therefore the polarization estimate for the entire sample was made using the portion of DESI's footprint with the highest quality imaging, the DES region. Although this results in a noisier measurement, only the average polarization of a sample affects the final RSD bias.\par

To estimate the polarization of the LRG redshift samples, we averaged the polarization estimates from the parent sample in bins of color and magnitude. LRGs in the redshift sample were assigned a polarization based on the average in their corresponding bin, and their total average is the polarization estimate of that sample. This was not done for the ELG sample, which were only used as tracers. A demonstration of this mapping can be seen in Figure \ref{fig:pol} and the results are also displayed in Table \ref{tab:redshift_comparison}. It is important to note that the polarization varies more across redshift bins than the IA signal, meaning that the redshift dependence of the final RSD bias is more dependent on survey selection than physical alignments.\par

\begin{table*}
\centering
\begin{tabular}{llllllcccc}
\hline
Tracer & z$_\text{min}$ & z$_\text{max}$ & N & $\sigma_{E1}^2$ &
$\epsilon_{\text{LOS}}$ & $\bar{\mathcal{E}}(0 < r_p < 20)$ & $\bar{L}(0 < r_p < 20)$& $\bar{\tau}(5 < r_p < 18)$ & $\bar{\xi}_\text{2, gI}(5 < s < 80)$\\ \hline

LRG & 0.4 & 0.6 & 529852 
& 0.046 & 2.3 $\times 10^{-3}$ & 
1.8 $\times 10^{-3}$ &
95.1 $h^{-1}{\rm Mpc}$& 
5.9 $\pm$ 0.5 $\times 10^{-3}$& 
0.044\\

LRG & 0.6 & 0.8 & 805181 
& 0.033 & 6.7 $\times 10^{-3}$ & 
2.1 $\times 10^{-3}$ &  
94.9 $h^{-1}{\rm Mpc}$& 
7.0 $\pm$ 0.2 $\times 10^{-2}$& 
0.22\\

LRG & 0.8 & 1.1 & 896150 
& 0.026 & 12.8 $\times 10^{-3}$ & 
1.8 $\times 10^{-2}$ & 
92.9 $h^{-1}{\rm Mpc}$& 
5.6 $\pm$ 0.2 $\times 10^{-2}$&
0.41 \\

ELG & 0.8 & 1.1 & 591687 
& 0.026 & 12.8 $\times 10^{-3}$ & 
1.9 $\times 10^{-3}$  & 
73.2 $h^{-1}{\rm Mpc}$& 
4.3 $\pm$ 0.3 $\times 10^{-2}$ &
0.34

\end{tabular}
\caption{Samples and values used to estimate the RSD bias for three LRG redshift bins and the LRGxELG cross-correlation. $r_p$ and $s$ are given in units of $h^{-1}{\rm Mpc}$. The tracer samples used in the top three rows were also used as the shape sample. The last row uses ELG tracers with LRG shapes. The table shows the redshift range and number $N$ of tracers used, and properties of the shape sample: the variance of the real component of ellipticities $\sigma_{E1}^2$ and the estimated selection-induced polarization of shapes along the LOS, $\epsilon_{\rm LOS}$. We do not include uncertainties for these columns as they have negligible statistical errors. The IA signal $\mathcal{E}(r_p)$ is measured as the ellipticity of shapes relative to the tracer sample. $L(r_p)$ is the effective LOS-distance that $\mathcal{E}(r_p)$ is averaged over. $\tau(r_p)$ is defined in Equation \ref{eq:tau} and is a combination of $\mathcal{E}(r_p)$, $L(r_p)$, and the power spectrum. These are functions of transverse separation, $r_p$ and are shown in this table as averages over the marked scales. The final column shows the average amplitude of the anisotropic clustering created by IA; the quadrupole of the correlation function without RSD effects. The full estimate of this final result along with the statistical error is shown in Figure \ref{fig:eRSD}.}
\label{tab:redshift_comparison}
\end{table*}

\begin{figure*}
\includegraphics[width=1\textwidth]{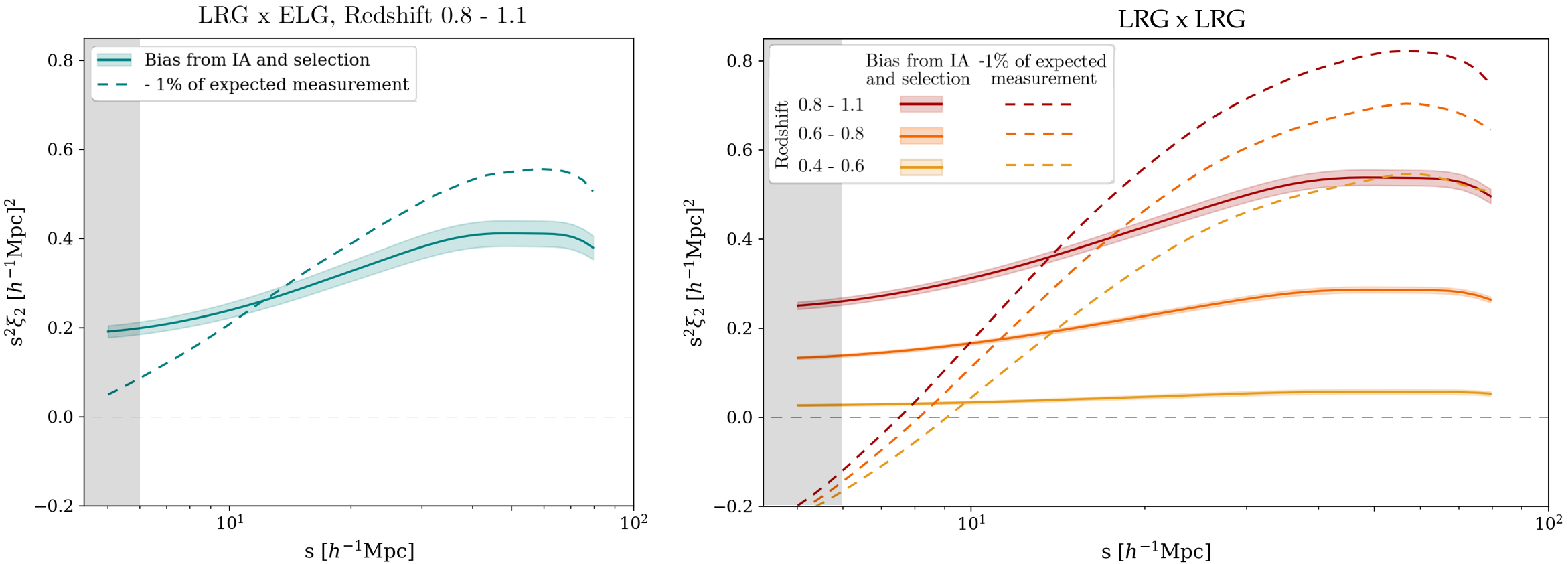}
\caption{The anisotropic clustering signal arising from tidal alignment and a selection bias, $\xi_\text{2, gI}$. Statistical errors are shown in the shaded bands, although the total errors are dominated by systematic effects (Section \ref{sec:conclusion}). Scales shaded in grey fall below the approximate validity of the non-linear IA model. For context, we have also plotted 1\% of the expected $\xi_2$ signal from RSD. This is well above DESI's error budget for measuring the growth rate of structure, which is 0.4-0.7\% for LRGs and ELGs combined.
Since the $\xi_2$ signal created by the growth of structure is opposite in sign to that created by IA, we have multiplied the RSD $\xi_2$ by -1 for an easier comparison. This plot demonstrates that IA will dampen DESI's RSD measurements to a degree that is a significant fraction of the error budget, particularly at higher redshifts. Incorporating these estimates into the $\xi_2$ measurement will mitigate this effect.}
\label{fig:eRSD}
\end{figure*}

\section{False RSD signature in DESI}\label{sec:result}

To estimate the RSD bias created by the combination of IA and the selection-induced polarization, we use a nonlinear tidal model adopted from  \cite{lamman_intrinsic_2023}. The full details can be found in this paper; we give only the results here. We have made minor notation changes for clarity.\par

The IA signal $\mathcal{E}$ is combined with the effective LOS-distance $L$, described in Section \ref{subsec:formalism}, and the nonlinear power spectrum $P$ as $\tau$:

\begin{equation}\label{eq:tau}
\tau = \frac{2 L(r_p) \mathcal{E}(r_p)}{ r_p \frac{d}{dr_p}\left[\frac{1}{r_p} \Psi\right]},
\end{equation}

\begin{equation}
\Psi(R) = \int \frac{K\;dK}{2\pi} \frac{P(K)}{K} J_1(KR)
\end{equation}

Here, $K$ is 2D Fourier Space and $J_1$ is the first Bessel function. $\tau$ is measured independently in each bin of transverse separation, $r_p$. The final variable used in our result, $\bar{\tau}$, is the average of these determinations with standard error. The transverse bins we used for determining $\tau$ were linear bins between $5-20h^{-1}$Mpc. Since these are relatively large scales, the change from a linear to nonlinear power spectrum had minimal effects on our final result, though it produced more consistent values of $\tau$ across the transverse bins.\par

The "false" signature this produces in the quadrupole of the correlation function, $\xi_2$, is
\begin{equation}\label{eq:xi_rsd}
\xi_\text{2, gI}(s)  = \epsilon_{\rm LOS} \frac{\bar{\tau}}{ 2\sigma_{E1}^2}
\int \frac{q^2 dq}{2\pi^2} P(q) j_2(qs).
\end{equation}
Here, $\epsilon_{\rm LOS}$ is the selection-induced shape polarization, $\sigma_{E1}^2$ is the variance of the shape parameter $\epsilon_+$ described in Section \ref{sec:desi_catalog}, $j_2$ is the second spherical Bessel function, and $s$ is 3D separation. The relations most relevant for this study can be summarized as
\begin{equation}
    \xi_\text{2, gI} \propto \epsilon_{\rm LRG} \frac{\bar{\tau}}{\sigma_{E1}^2} \propto \epsilon_{\rm LRG} \frac{L \mathcal{E}}{\sigma_{E1}^2}.
\end{equation}
Note that, through cancellation, this result is independent of the amplitude of the power spectrum and galaxy bias, $b$. This is because $\xi_\text{2, gI}$ arises from the correlation of the galaxy density field and the selection-induced shape polarization, the latter of which is independent of bias. It does depend on the projected correlation function $w_p$ through $L$. Also, since the IA signal only affects $\xi_\text{2, gI}$ through $\bar{\tau}$, which can be determined in transverse bins independently, we can forecast $\xi_\text{2, gI}$ beyond the projected scales used to measure $\mathcal{E}$.

The variables measured for this estimate are listed in Table \ref{tab:redshift_comparison} and the final quadrupole signature for all our samples is shown in Figure \ref{fig:eRSD}. To provide context for this signal, we estimate the total quadrupole signatures $\xi_2$ expected for these galaxy samples. They are based on HOD catalogs made with the A\textsc{bacus}S\textsc{ummit} simulations \citep{hadzhiyska_compaso_2021, maksimova_abacussummit_2021, yuan_inprep}, and scaled with measurements from DESI's Survey Validation \citep{desi_collaboration_validation_2023}. Figure \ref{fig:eRSD} shows 1\% of these estimates, which is well above DESI's total error budget for measuring $\xi_2$. Since the $\xi_2$ signal created by the growth of structure is opposite in sign to that created by IA, we have also multiplied this 1\% line by -1 for a clear comparison.
On the scales used to measure $f\sigma_8$ (10 $< s <$ 80 $h^{-1}$Mpc), $\xi_2$ for LRGs will be dampened by around 0.15\% between redshifts 0.4-0.6, 0.53\% between redshifts 0.6-0.8, and 0.80\% between redshifts 0.8-1.1. $\xi_2$ as measured by LRG x ELG cross-correlations will be biased by around 0.83\% between redshifts of 0.8-1.1. These are a significant fraction of DESI's forecasted error on $f\sigma_8$ (0.4-0.7\%), which is dominated by the error $\xi_2$.\par

We used a Nonlinear Alignment model, which has shown to be valid down to 6 $h^{-1}$Mpc for LRGs \citep{singh_intrinsic_2015}. In principle our estimate can be extended down to the scales of the Fingers of God effect, where peculiar velocities of galaxies create a "smearing" along the LOS, as opposed to the "squashing" created by structure growth \citep{jackson_critique_1972}. Here, the sign of $\xi_2$ switches and this bias will result in an enhancement of the signal. However, as nonlinear effects become more apparent here and this effect is less relevant for DESI's main science goals, it is most valuable to interpret the bias on large scales.\par

\section{Conclusion}\label{sec:conclusion}

We measure the tidal alignment of LRGs with DESI Year 1 redshifts, using both LRG and ELG tracers. We also estimate a redshift-dependent polarization in LRG orientations relative to the LOS which arises from an aperture-based target selection. Using a nonlinear tidal model, we calculate the signal this will create in DESI's measurements of the quadrupole of the correlation function. It ranges from 0.2-1.1\% of the quadrupole signal created by RSD. This is a significant fraction of DESI's full-survey error budget of around 0.4-0.7\% for measuring the growth rate with LRGs and ELGs combined.\par

The RSD bias is over five times larger in the highest redshift sample, $0.8<z<1.1$, than the lowest, $0.4<z<0.6$. This is partially due to a stronger alignment signal, but mostly due to the selection effect. Galaxies at higher redshifts are redder and fainter, falling closer to the target selection cuts. Therefore their orientation has a stronger influence on whether or not they pass the aperture magnitude cut and the sample has a stronger orientation polarization. If uncorrected for, this redshift-dependent bias will suppress measurements of the growth rate at all scales, but especially higher redshifts. Therefore it will also bias determinations of how the growth rate evolves, a critical estimator for constraining cosmological models \citep{kazantzidis_8_2021}. Our estimates of the quadrupole signature created by IA presented here can be used to adjust initial estimates of $xi_2$ and mitigate this bias.\par


These results agree with the previous work in \cite{lamman_ia_2023}, which measured the IA signal using photometric redshifts and estimated that it will produce around a 0.5\% decrease on measurements of $\xi_2$ with the full LRG sample. While large, upcoming photometric surveys can provide constraints on IA, for this effect it's most important to understand the IA of our particular sample. Additionally, redshifts are necessary to make clean distant cuts to explore redshift dependence in both IA and the shape polarization.

The largest uncertainty in our final results comes from systematic effects in the estimate of the selection-induced shape polarization. This is sensitive to assumptions in the light profiles used for mock selection and the underlying triaxial distribution of shapes, which is based on SDSS imaging \citep{padilla_shapes_2008} and not in clear agreement with comparable galaxies in hydrodynamic simulations \citep{bassett_prospects_2019}. This could be significantly improved with a large imaging survey such as the Dark Energy Survey or the upcoming Legacy Survey of Space and Time \citep{gatti_dark_2021, ivezic_lsst_2019}. The methods of imaging and shape fits have a known impact on the inferred intrinsic shapes of galaxies \citep{georgiou_dependence_2019, macmahon_intrinsic_2023}.\par

While we do not expect a significant RSD bias from the shape alignment of ELGs, their spins are known to correlate with the tidal field \citep{Lee_intrinsic_2011}. Their shapes could also be biased in DESI's sample, as spectroscopic quality depends upon disk orientation \citep{hirata_tidal_2009}. This bias could be explored in DESI through correlations between the quadrupole of the correlation function and fundamental plane residuals \citep{singh_fundamental_2021}.\par

The remaining four years of DESI's main survey will produce millions more spectroscopic redshifts, allowing us to refine IA measurements and their redshift dependence. These will also produce higher precision RSD measurements, necessitating the need to incorporate the anisotropic clustering effect caused by IA.

\section*{Acknowledgements}
CL thanks the DESI internal reviewers of this paper, Mustapha Ishak and Benjamin Joachimi, for through feedback. CL also thanks Jonathan Blazek for several helpful discussions.

This material is based upon work supported by the National Science Foundation Graduate Research Fellowship under Grant No. DGE1745303, the U.S.\ Department of Energy under grant DE-SC0013718, NASA under ROSES grant 12-EUCLID12-0004, and the Simons Foundation.

This material is based upon work supported by the U.S. Department of Energy (DOE), Office of Science, Office of High-Energy Physics, under Contract No. DE–AC02–05CH11231, and by the National Energy Research Scientific Computing Center, a DOE Office of Science User Facility under the same contract. Additional support for DESI was provided by the U.S. National Science Foundation (NSF), Division of Astronomical Sciences under Contract No. AST-0950945 to the NSF’s National Optical-Infrared Astronomy Research Laboratory; the Science and Technology Facilities Council of the United Kingdom; the Gordon and Betty Moore Foundation; the Heising-Simons Foundation; the French Alternative Energies and Atomic Energy Commission (CEA); the National Council of Science and Technology of Mexico (CONACYT); the Ministry of Science and Innovation of Spain (MICINN), and by the DESI Member Institutions: \url{https://www.desi.lbl.gov/collaborating-institutions}.

The DESI Legacy Imaging Surveys consist of three individual and complementary projects: the Dark Energy Camera Legacy Survey (DECaLS), the Beijing-Arizona Sky Survey (BASS), and the Mayall $z$-band Legacy Survey (MzLS). DECaLS, BASS and MzLS together include data obtained, respectively, at the Blanco telescope, Cerro Tololo Inter-American Observatory, NSF’s NOIRLab; the Bok telescope, Steward Observatory, University of Arizona; and the Mayall telescope, Kitt Peak National Observatory, NOIRLab. NOIRLab is operated by the Association of Universities for Research in Astronomy (AURA) under a cooperative agreement with the National Science Foundation. Pipeline processing and analyses of the data were supported by NOIRLab and the Lawrence Berkeley National Laboratory. Legacy Surveys also uses data products from the Near-Earth Object Wide-field Infrared Survey Explorer (NEOWISE), a project of the Jet Propulsion Laboratory/California Institute of Technology, funded by the National Aeronautics and Space Administration. Legacy Surveys was supported by: the Director, Office of Science, Office of High Energy Physics of the U.S. Department of Energy; the National Energy Research Scientific Computing Center, a DOE Office of Science User Facility; the U.S. National Science Foundation, Division of Astronomical Sciences; the National Astronomical Observatories of China, the Chinese Academy of Sciences and the Chinese National Natural Science Foundation. LBNL is managed by the Regents of the University of California under contract to the U.S. Department of Energy. The complete acknowledgments can be found at \url{https://www.legacysurvey.org/}.

Any opinions, findings, and conclusions or recommendations expressed in this material are those of the author(s) and do not necessarily reflect the views of the U. S. National Science Foundation, the U. S. Department of Energy, or any of the listed funding agencies.

The authors are honored to be permitted to conduct scientific research on Iolkam Du’ag (Kitt Peak), a mountain with particular significance to the Tohono O’odham Nation.

\section*{Data Availability}
The DESI Legacy Imaging Survey is publicly available at \href{https://www.legacysurvey.org/}{legacysurvey.org} and DESI's Early Data Release is publicly available at \href{https://data.desi.lbl.gov/doc/releases/edr/}{data.desi.lbl.gov/doc/releases/edr/}. Iron covers the DESI Year 1 sample and will be publicly released as part of DESI Data Release 1 (DR1). AbacusSummit simulations are publicly available at \href{https://abacusnbody.org/}{abacusnbody.org}. Code for projecting ellipsoids and generating light profiles can be found at \href{https://github.com/cmlamman/ellipse_alignment}{github.com/cmlamman/ellipse\_alignment}.\par
Data plotted in this paper are available at \href{https://zenodo.org/uploads/10162040}{zenodo.org/uploads/10162040}.



\bibliographystyle{mnras}
\bibliography{references} 





\bsp	
\label{lastpage}
\end{document}